\documentstyle[12pt,epsfig]{article}
\def\ifig#1#2#3#4{\begin{figure}[hbtp]
 \begin{center}\leavevmode\epsfig{file=#1,height=#2}
 \caption[#3]{#4}\end{center}\end{figure}}
\def\figlr#1#2#3#4#5{
  \begin{figure}[hbtp]\begin{center}\leavevmode
\epsfig{file=#1,height=#3}\epsfig{file=#2,height=#3}\hfill\hfill
  \caption[#4]{#5}\end{center}\end{figure}}

\title{ Stability of Bubble Nuclei through Shell-Effects
\footnote{We dedicate this paper to Prof. Richard Lemmer on the occasion
          of his 65$^{th}$ birthday.}}

\author{     Klaus Dietrich and Krzysztof Pomorski
    \thanks{On leave on absence from University M.C.S. in Lublin}\\
      Technische Universit\"at M\"unchen, Garching, Germany}

\date{}

\begin{document}
\maketitle
          
\begin{abstract}

We investigate the shell structure of bubble nuclei
in simple phenomenological shell models and study
their binding energy as a~function of the radii and of the
number of neutron and protons using Strutinsky's method.
Shell effects come about, on the one hand,
by the high degeneracy of levels with large angular
momentum and, on the other, by the big energy gaps between
states with a~different number of radial nodes.
Shell energies down to -40~MeV are shown to occur for
certain magic nuclei. Estimates demonstrate
that the calculated shell effects for certain magic 
numbers of constituents are probably large enough to
produce stability against fission, $\alpha$-, and $\beta$-decay.
No bubble solutions are found for mass number $A \leq 450$.\\

PACS numberes: 21.10.Sf,24.10.Nz,47.20.Dr,47.55.DZ
\end{abstract}

\section{Introduction}

The possibility that nuclei could exist in the form of a (spherical)
bubble or in the form of a toroid has been pointed out long ago [1,2].
Within the liquid drop model (LDM) nuclei of these shapes turn out
to be unstable with respect to deformations. Shell--effects may, however,
stabilize such nuclei against deformation. 

In a series of papers [3], C.Y. Wong investigated shell--effects
for toroidal and bubble-shaped nuclei using Strutinsky's shell correction
method. He restricted his attention to known nuclei near the valley of 
$\beta$--stability  and found that for certain doubly magic nuclei
($^{200}_{\,80}Hg_{120},^{138}_{\,58}Ce_{80}$) spherical bubble
solutions with a very small inner radius (ratio of inner to outer 
radius $\approx $0.07) turned out to be the ground state.
Indications that bubble solutions might exist were also found in mean field
calculations [4] and for stellar matter at finite temperature [5].
More recently, Moretto et al. [6] showed in a classical model that
LD--bubbles at finite temperature may be stabilized by an internal 
vapor pressure.

In the present paper, we study shell effects for nuclear bubbles in a
broad range of neutron ($N$) and proton ($Z$) numbers extending
considerably beyond the known nuclei. As C.Y. Wong we make use
of Strutinsky's shell correction method [7].
We show that the shell energy may become as large as
-40~MeV for certain magic numbers of the nuclear constituents
and that nuclear bubbles may thus become stable or very
long-lived against fission and other decay modes.

\section{Theory}

In Strutinsky's method [7], the total binding energy $E$ of
a~nucleus (neutron number $N$, proton number $Z$, nucleon number
$A$) of given shape is given as a sum of the liquid drop (LD)
energy $E_{\rm LD}(N,Z)$ and the shell correction energy
$\delta E_{\rm shell}$
\begin{equation}
 E(N,Z) = E_{\rm LD}(N,Z) + \delta E_{\rm shell}(N,Z)\,\,.
\end{equation}
The shell correction energy has the well-known form [7]
\begin{equation}
 \delta E_{\rm shell} = \sum_\nu e_\nu \, \delta n_\nu\,,
\end{equation}
where $e_\nu$ are the single particle energies and $\delta n_\nu$
is the difference between the occupation numbers $n_\nu =
\theta_0(e_F - e_\nu)$ in the shell-model ground state and
a~smooth distribution $\bar n_\nu$
\begin{equation}
 \delta n_\nu = n_\nu - \bar n_\nu\,.
\end{equation}
The smooth occupation pattern is defined in the usual way as
a~functional of a~smooth level distribution [7].

We use the Strutinsky method to study the total energy and
especially the shell correction energy of spherical nuclear
bubbles. 
The single particle energies $e_\nu$ as well as the LD energy
depend on the inner ($R_2$) and outer ($R_1$) radius of the
bubble nucleus. Adopting the conventional saturation condition
that the volume of the LD remains constant independently of its
shape, the two radii are related by the condition
\begin{equation}
 R^3_1 - R^3_2 = R^3_0\,,
\end{equation}
where $R_0 = r_0 A^{1/3}$ is the radius of a compact spherical
nucleus of the same mass. We describe the shape of the bubble
nucleus either by the dimensionless radii
\begin{equation}
 v_{1(2)}: = {R_{1(2)}\over R_0}
\end{equation}
or by the ratio $f$ between the volume of the hole and the
volume of the entire bubble
\begin{equation}
 f: = {R^3_2\over R^3_1}\,.
\end{equation}
The difference $\Delta E_{\rm LD}$ between the energy of the
LD-bubble and the energy of the corresponding compact spherical
LD is given by
\begin{equation}
\Delta E_{\rm LD} = E_S(R_1,R_2) + E_{\rm Cb}(R_1,R_2) -
   E_S(R_0) - E_{\rm Cb}(R_0)\,,
\end{equation}
where $E_S$ and $E_{\rm Cb}$ are the surface and Coulomb
energies of the bubble and the compact configuration, resp.
Measuring $\Delta E_{\rm LD}$ in units of the surface energy
$E_S(R_0)$ and substituting the explicit form of the different
energies, we find
\begin{equation}
 {\cal E}:  =  {\Delta E_{\rm LD} \over  E_S(R_0)} =
   v^2_1(v_2) + v^2_2 - 1 + 2X_0 \left[v^5_1(v_2) + {3\over 2}
  v^5_2 - {5\over 2}v^3_2 v^2_1(v_2) - 1\right]\,,
\end{equation}
where 
\begin{equation}
  v_1(v_2): = [1 + v_2^3]^{{1\over 3}}
\end{equation}
and $X_0$ is the conventional fissility parameter 
\begin{equation}
X_0 \equiv {E_{\rm Cb}(R_0)\over 2 E_S(R_0)} = {Z^2/A\over
   (Z^2/A)_{\rm crit}}
\end{equation}
with
\begin{equation}
\left(Z^2/A\right)_{\rm crit}: = {40\pi\sigma r^3_0\over 3e^2_0}\,.
\end{equation}
The quantities $\sigma$, $r_0$, and $e_0$ are the surface
constant, the radius parameter, and the elementary charge, resp..
Beside the trivial solution $v_2 = 0$ (compact spherical
nucleus) the condition of stationarity
\begin{equation}
 {\partial{\cal E}\over \partial v_2} = 0
\end{equation}
has real solutions $v_2(X_0)$ only for $X_0 > 2.02$. They
represent a~multivalued function which is shown in the l.h.s. of Fig. 1
for a limited range of values $v_2(X_0)$. For
given $X_0 > 2.02$ the smallest solution $v_2(X_0) \leq 0.4$
corresponds to the maximum of the potential barrier, whereas the
next solution in magnitude $v_2(X_0) > 0.4$ corresponds to
a~minimum of the energy ${\cal E}$ for a bubble of the reduced
radius $v_2(X_0)$. 

On the r.h.s. of Fig. 1 we show the energy change ${\cal E}$ by bubble
formation as a function of $v_2$. For $X_0 > 2.2$, the bubble
solution is seen to correspond to a lower energy than the one of
a compact sphere ($v_2 = 0$). The barrier between the bubble
solution and the compact spherical one occurs for reduced inner
radii in the interval $0 \leq v_2 \leq 0.4$. 
In units of $E_S(R_0)$ the barrier heights are seen to be at
most at ${\cal E} = 0.025$. The unit $E_S = 4\pi\sigma R^2_0$
is, however, pretty large. For $A=580$, $Z=280$ and reasonable
values of the parameters $r_0$ and $\sigma$, we have $E_S
=1291$~MeV and $X_0=2.62$. The corresponding energies of the
barrier and the minimum are $\sim 20$~MeV and $\sim
-140$~MeV. 
It is not meaningful to pursue the solutions $v_2$ of Eq.
(12) to values of $v_2$ and $X_0$ much larger than the ones
given in Fig. 1, because for increasing $X_0$ the charge density
of the bubble turns out to be unphysically large and the
diameter $d= R_1-R_2$ of the bubble layer becomes unphysically
small. 

We emphasize that the bubble solutions obtained in the LDM are
not stable with respect to deformations [3] in the same way as
the compact spherical liquid drops are not stable agains fission
for fissibility parameters $X_0 > 1$.
Nevertheless, stability can in principle be produced by shell
effects. 

As two extreme and simple cases of nuclear single particle
potentials we consider the shell effects in an infinite square
well and in a harmonic oscillator:
\begin{equation}
 V(r) = \left\{\begin{array}{ll}
 -V_0 & \mbox{for}~~R_2 < r < R_1 \\
  +\infty & \mbox{otherwise}
\end{array}\right.
\end{equation}
\begin{equation}
  V(r) = -V_0 + {M\omega^2\over 2} (r - \bar R)^2\,. 
\end{equation}
The depth $V_0 > 0$ has no influence on the shell correction
energy and can thus be put equal to zero. The center of the
oscillator potential is chosen to be
\begin{equation}
 \bar R = {R_1 + R_2\over 2}\,.
\end{equation}
The oscillator frequency $\omega$ is chosen in such a way that
the RMS deviation from the sphere of radius $\bar R$ is the same
when calculated with the shell model wave function and with the
LD-density 
\begin{equation}
\langle(r - \bar R)^2\rangle_{\rm SM} = \langle(r - \bar R)\rangle_{\rm LD}\,.
\end{equation}

We have to add a spin--orbit term to the central potential (13) or (14).
We use the conventional form in the Skyrme approach (see Eq. (5.103) in 
Ref. [8])
\begin{equation}
 \widehat V_{\rm SO}  = \tilde V_{\rm SO}(r) \, \widehat{\vec l}
 \cdot \widehat{\vec s}\,\,,
\end{equation}
\begin{equation}
\tilde V_{\rm SO}(r) = {3\over 2}\, W_0 \, {1\over r}
   {\partial\rho(r)\over \partial r}\,\,,
\end{equation}
$\rho(r)$ is the nuclear density distribution. We choose the value 
$W_0 = 120$~MeV~fm$^5$ given in Ref. [8] and neglect the isospin
dependence which is recently under debate [9].

For the total density $\rho(r)$ which appears in the expression
for the spin-orbit potential $\tilde V_{\rm SO}(r)$ (Eq.
(18)) we used the following ansatz:
\begin{equation}
\rho(r) = \rho_0 \left[1 - 2 \left(\frac{r- \bar R}{R_1-R_2}\right)^2 \right]
\end{equation}
where $\rho_0$=0.17 fm$^{-3}$ and $\bar R$ is given by Eq. (15). 
For simplicity we have assumed that the proton and neutron densities 
$\rho_p$ and $\rho_n$ are proportional to $Z$ or $N$ respectively.

It is seen from (18) that the sign of the spin-orbit potential
is opposite in the inner and outer surface region of the bubble
nucleus. Consequently, the magnitude of the spin-orbit splitting
is smaller for bubble nuclei than for normal ones. We, therefore,
treat the spin-orbit term in perturbation theory.
The eigenenergies $e_{nlj}$ of the s.p. Hamiltonian including the
spin--orbit potential (17) and the unperturbed eigenvalues $\varepsilon_{nl}$
are related by the equation
\begin{equation}
 e_{nlj} = \varepsilon_{nl} + \langle\psi_{nljm}|\tilde V_{\rm SO}|
 \psi_{nljm}\rangle \cdot \left(j(j+1) - l(l+1) - {3\over 4}\right)\,\,,
\end{equation}
where ($n-1$) represents the number of radial nodes (not counting
zeros at $r=0,\infty$) and $l,j,m$ are the orbital and total angular
momentum and its projection, resp.. The mean value $\langle\psi_{nljm}|
\tilde V_{\rm SO}| \psi_{nljm}\rangle$ depends only on the unperturbed
s.p. density
\begin{equation}
\rho_{nl}(r) = \psi^+_{nljm}(\vec r) \psi_{nljm}(\vec r) = 
 {u^2_{nl}(r)\over r^2} \,\,.
\end{equation}
The functions $u_{nl}(r)$ satisfy the radial Schr\"odinger equation with
eigenvalue $\varepsilon_{nl}$. For the infinite square well the 
eigenvalues and eigenfunctions are obtained by numerically satisfying
the boundary conditions and for the harmonic oscillator (14) we used 
the WKB approximation.
\section{Results and Discussion}

We complement the results obtained in the pure LDM (see Fig. 1)
by Fig. 2 which shows lines of equal LD-binding energy per
particle $\left({E_{\rm LD}(N,Z)/A}\right)$ as a function of $N$
and $Z$. We note that the volume term of the LDM is present in
this figure since we do not subtract the energy of the compact
spherical LD. The LD parameters are taken from Ref. \cite{My66}.
Each point on the equipotential lines corresponds
to a spherical bubble solution as given by the LDM, which is
unstable with respect to deformations.
It should be noted that the largest energy gains per
particle occur for nucleon numbers 1200 $\leq A \leq 2000$ and the
corresponding proton numbers $325 \leq Z \leq 400$.  Some
of the isobaric lines are cut twice by the same equi--energy line.
If the binding energy per particle in between these 2 points is smaller,
a $\beta$-stable isobar lies somewhere on this section. Of
course, this consideration is thwarted by the fact, that the
LD-bubbles are all unstable against fission.  In the LDM, the
(spherical) bubble solutions are saddle points, not minima.

The LD results (and consequently also the result on the total
binding energy) depend sensitively on the value of the surface
constant $\sigma$. This surface constant contains a~poorly known
isospin-dependent part. In our calculations we adopted the
isospin dependence given by Myers and Swiatecki \cite{My66} who write
the surface energy of a~spherical LD in the form
\begin{equation}
 E_S(R_0) = 17.9439~{\rm MeV} \cdot A^{2/3} \left[1 - 1.7826 
 \left({N-Z\over A}\right)^2\right]
\end{equation}

In Fig. 3 we show the spectrum of single particle levels for the
shifted infinite square well with a~spin-orbit term. The levels
are shown as a function of the hole fraction parameter $f$ (see
Eq. (6)).  As $f$ increases, the diameter $d = R_2 - R_1$ of
the bubble layer decreases. Increasing the number of radial
nodes for given orbital angular momentum $l$ thus costs an
energy which rises steeply as a function of $f$. Augmenting a
given $l$-value by 1 for given $n$ and given parameter $f$
implies an increase of the centrifugal energy, which is the
larger, the larger the $l$-value. Magic numbers come about by
the interplay between levels $e_{nlj}$ with $n \geq 2$, which
rise rapidly as a function of $f$ and levels $e_{1lj}$, which
depend more gently on $f$ with a tendency to decrease with $f$
due to the diminishing centrifugal energy.

In Fig. 4, we display a corresponding level spectrum for the
harmonic oscillator potential. The general trends are the same as
for the infinite square well, but the magic numbers for the same
values $f$ differ for the square well and oscillator. This is
not surprising. A Saxon-Woods form centered around $\bar R$
would be more realistic and the magic numbers for this choice
would be expected to lie between the limits given by the
infinite square well and the oscillator.

For both the potentials we observe that the spin-orbit splitting
is often reversed as compared to the case of normal nuclei.  The
reason was already given in Section 2.  The shell correction
energy $\delta E_{\rm shell}$ is shown as a function of $N$ (or
$Z$) at a given value of $f=0.28$ for the infinite square well
in Fig. 5 and for the harmonic oscillator in Fig. 6. Please note
the different energy scales in case of the square well and the
oscillator. The eigenenergies of the square well scale with
$A^{-2/3}$ which is taken into account by the energy unit. No
such simple scaling property exists for the oscillator. It is
seen from the Figs. 5 and 6 that the shell energy may produce
energy gains up to -20~MeV for one sort of particles. Of course,
double magic shell closures can only occur, if the two magic
numbers correspond to the same $f$-values. 

In Fig. 7 we show lines of constant shell correction energy
$\delta E_{\rm shell}$ in the $(N,Z)$-plane. Each point
corresponds to a minimum of the total energy as a function of
$f$. One should notice that for this case of relatively light
bubbles, the shell effects are especially large for almost
symmetrical nuclear composition.

In Fig. 8 we display lines of equal energy gain by formation
of a bubble. As a reference we use the energy of a spherical LD
of the same numbers of neutrons and protons. It is seen that the
gain in binding energy may amount to several 100~MeV.

We have still to deal with the crucial question of stability of
the bubble solutions with respect to shape deformations.
Calculating Strutinsky's shell correction energy (2) for a
deformed bubble nucleus implies that we find the eigenvalues of
a Schr\"odinger equation in 3 dimensions or at least (for axial
symmetry) in 2 dimensions. Without solving this technical
problem we may obtain a valuable insight by describing the deformation
dependence of the shell correction energy as suggested
by Myers and Swiatecki \cite{Sw66}.
In Fig. 9 we display the dramatic effect of the shell
correction on the binding energy of a bubble nucleus as a
function of the quadrupole deformation. The total LD binding energy
of the spherical bubble nucleus is put equal to zero. The
LD-part of the energy decreases monotonically as $|\beta_2|$
increases. Adding the shell energy $\delta E_{\rm shell }$ with
Swiatecki's $\beta_2$-dependence produces a valley of about
-30~MeV depth. The barriers on both sides of the minimum are
about at an energy of -3.9~MeV which leaves us with a barrier
height of about 25~MeV. This order of magnitude of the fission
barrier implies an almost vanishing probability for spontaneous
fission. 

What about the other decay modes? $\beta$-decay will imply that
the system moves along isobaric lines in the $(N,Z)$-plot
towards lower energy. Fig. 8 is particularly
instructive in this respect: If a given isobaric straight line
cuts a line of constant binding energy twice and if, at the same
time, the binding energy between the two points of intersection
is lower, a $\beta$-stable nuclear bubble lies in between.
In Fig. 8 this happens to be the case for $A\geq 600$.
If, on the other hand, the energy keeps on lowering
along a given isobaric direction, the $\beta$-decays may 
finally lead to a nuclear composition where bubbles cease to exist.
Thus there are cases where bubble nuclei may disintegrate by
a series of $\beta$-decays and others where the $\beta$-decays
make them approach a stable composition.

The $\alpha$-decay, which limits the lifetime of the presently
known superheavy nuclei, certainly may also limit the lifetime
of bubble nuclei. The penetrability of the Coulomb barrier for
an $\alpha$-particle depends exponentially on the Coulomb
potential at $r = R_1$, which has the value $2Ze_0/R_1$.
The higher the Coulomb potential, the lower the $\alpha$-decay
probability. For $A =700$, $Z=270$, and $R_1=11$~fm one finds 
$2Ze^2_0/R_1 \approx $~70~MeV. We estimate the $Q$-value to
be
$$
Q_\alpha: = E_{\rm tot}(Z,N) - [E_{\rm tot}(Z - 2, N- 2) -
   28~{\rm MeV}]\, \approx \, 20 MeV \,\,.
$$
Thus the $\alpha$-particle has to tunnel through a Coulomb
barrier whose top is $\sim$50~MeV above the tunneling energy.
Empirically, the $\alpha$-decay of heavy actinide nuclei is
found to be vanishingly small whenever the emitted
$\alpha$-particle has an energy $Q_\alpha \leq 4$~MeV, i.e. if
the energy difference 
$$
\left({2Z_{\rm act.} e_0^2\over R_0({\rm act.})} - Q_\alpha\right) \geq 
\left({2Z_{\rm act.} e_0^2\over R_0({\rm act.})}- 4\right) \approx 30 MeV\,\,.
$$ 
This estimate implies that the
$\alpha$-decay probability of typical bubble nuclei is very
small due to the especially large Coulomb barrier.
Of course, much more careful investigations of the decay modes
have to be performed. We may, however, safely say that certain
bubble nuclei are expected to have a practically infinite
lifetime.

It is intriguing to imagine that stable or at least very
long-lived bubble nuclei may exist. Their properties would be a
fascinating subject of research. Unfortunately, the masses and
charges of the best candidates for bubble structure are so high
that there is no hope to produce them. More careful work with
the full Hartree-Bogoliubov theory is of course
necessary, especially for determining the lower limits of mass
and charge numbers of these objects. Even if bubble nuclei can
never be made in an earthly laboratory, they might play a role
in neutron stars [5].

Finally, there may be bubble structures for mesoscopic systems
consisting of some 1000 atoms. This was already conjectured in
Ref. [6]. We believe that it may be also of interest to investigate
whether shell effects favour the bubble topology for certain
mesoscopic systems.

\bigskip
\noindent
{\bf Acknowledgments}

\medskip

K. Pomorski is grateful to the DFG for granting a~guest
professor position which  greatly facilitated the present 
research. Prior to the 1$^{\rm st}$ April 1997, this research was
also modestly supported by the BMBF which is gratefully
acknowledged. This work was also partially financed by the Polish Committee
of Scientific Research under contract No. 2P03B~049~09.
K.D. is grateful for numerous stimulating e-mail
messages by M. Weiss.

\newpage
{\bf Figures captions.}

\begin{enumerate}

\item The condition (12) defines a function $v_2(X_0)$ which is shown in 
      the l.h.s. of the figure. The r.h.s. shows the difference 
      ${\cal E}= \Delta E_{\rm LD}/E_S(R_0)$ of the LD energy of 
      a bubble of reduced inner radius $v_2$ and the LD energy of a compact 
      spherical nucleus of radius $R_0$.

\item Lines of constant LD--binding energy per particle as a function of 
      ($N,Z$). Each point on the lines corresponds to a bubble solution 
      within the LDM. The straight lines in the plot represent isobars.

\item Level scheme as a function of the hole fraction $f=(R_2/R_1)^3$
      for the infinite square well plus the spin orbit term (Eqn. 17-19).
      For 3 values of $f$ (0.12, 0.24, 0.28) numbers are written just 
      above certain s.p. energies $e_\nu$ whenever the distance
      ($e_{\nu +1} - e_\nu$) exceeds 1.5 energy units.
      Following the conventional  spectroscopic notation the l-values
      of the levels are given by the letters of the alphabet. The energy 
      unit used takes into account that the eigenvalues of the infinite 
      square well scale with $A^{-2/3}$.

\item As Fig. 5, but for the harmonic oscillator potential. The oscillator 
      frequency $\omega$ was determined as a function of $f$ for $A$=500
      using relation (16).

\item Shell correction energy as a function of $Z$ (or $N$) for $f$=0.28
      for the infinite square well.

\item As in Fig. 7, but for the harmonic oscillator potential. The 
      frequency of the oscillator was determined by the relation (16) for
      the mass number $A$=500.

\item Lines of constant shell energy $\delta E_{shell}$ in the ($N,Z$)
      plane. $\delta E_{shell}$ was calculated for the infinite square
      well. Each point on the equi-energy lines is evaluated for the
      hole fraction parameter $f$ corresponding to the minimum of the
      total energy.

\item Lines of constant energy gain ($\Delta E_{\rm LD} + \delta E_{shell}$)
      with respect to the energy of a compact spherical LD. The shell 
      correction energy was calculated  with the infinite square well 
      potential.

\item LD--energy (dashed line), shell correction energy (dotted line) and 
      total energy (solid line) as a function of the quadrupole deformation 
      $\beta_2$ of the outer bubble surface $S_1$. The liquid drop energy
      was minimized with respect of the deformations $\beta_4$ and $\beta_6$
      of the outer surface and $\beta_2$ to $\beta_6$ of the inner surface.
      The deformation dependence of the shell correction was taken from
      Ref. \cite{Sw66}.

\end{enumerate}

\figlr{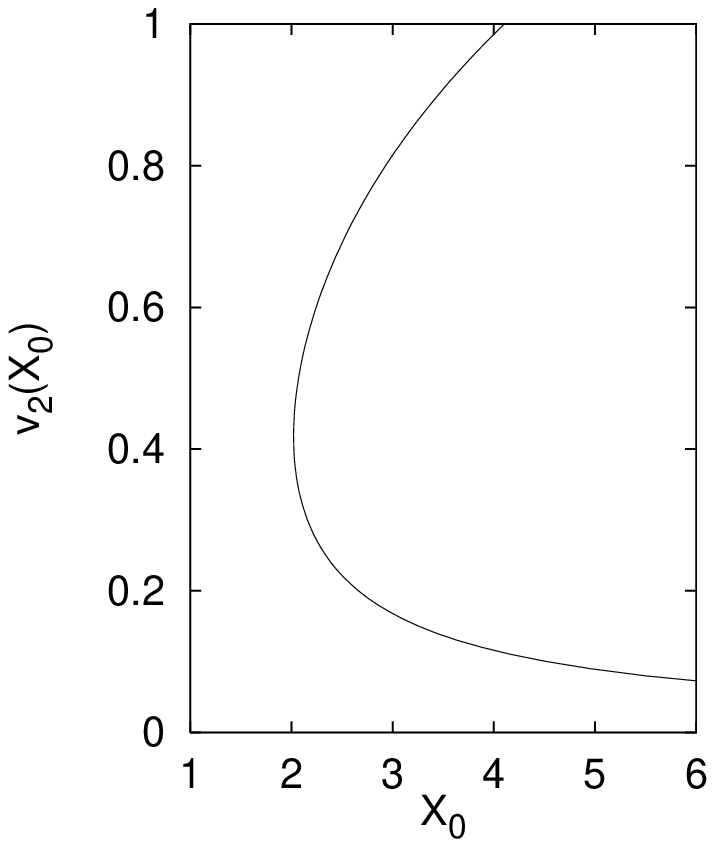}{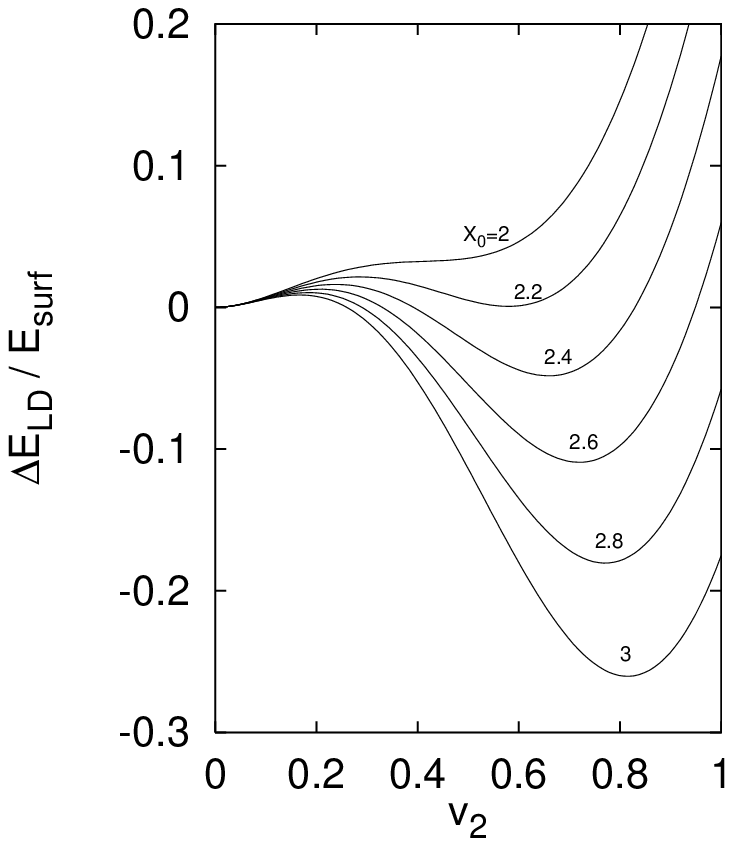}{60mm}{1}{} 
\pagebreak[5]
\ifig{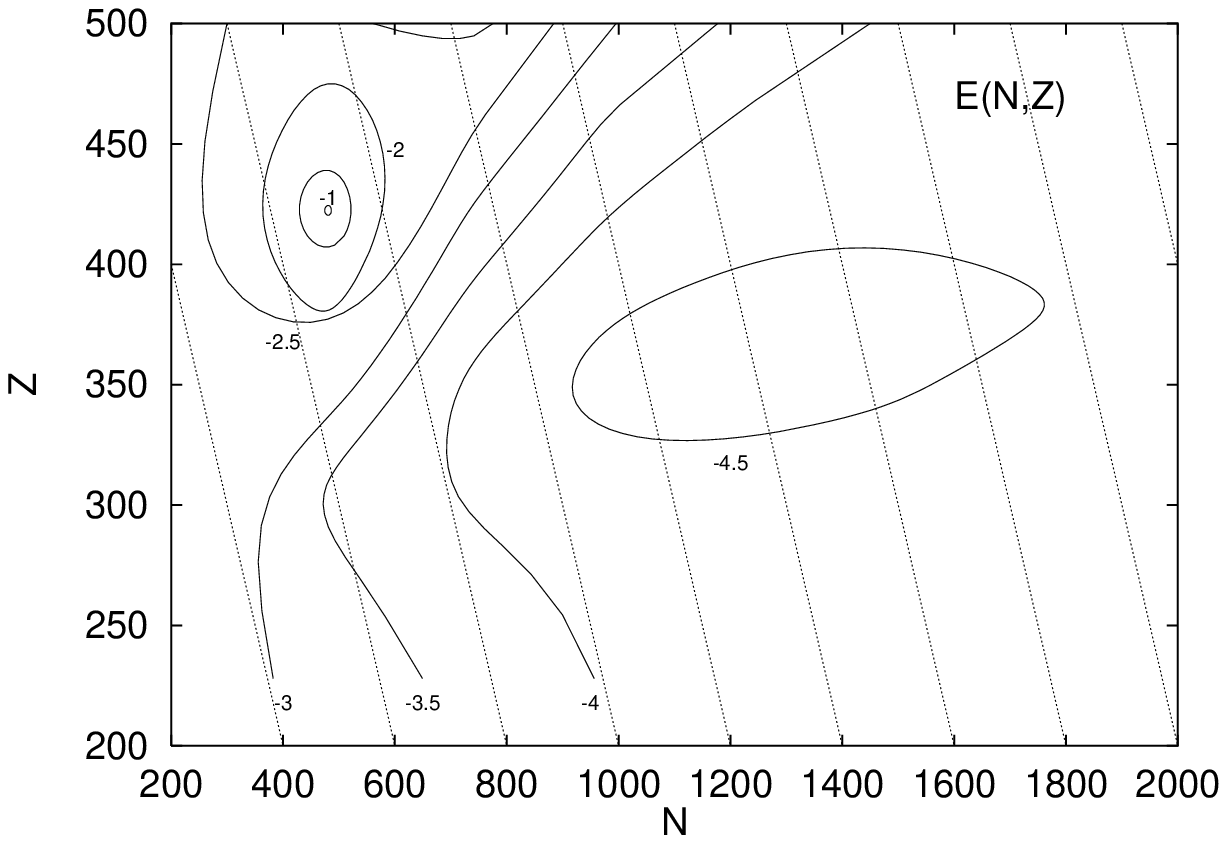}{100mm}{2}{}
\pagebreak[5]
\ifig{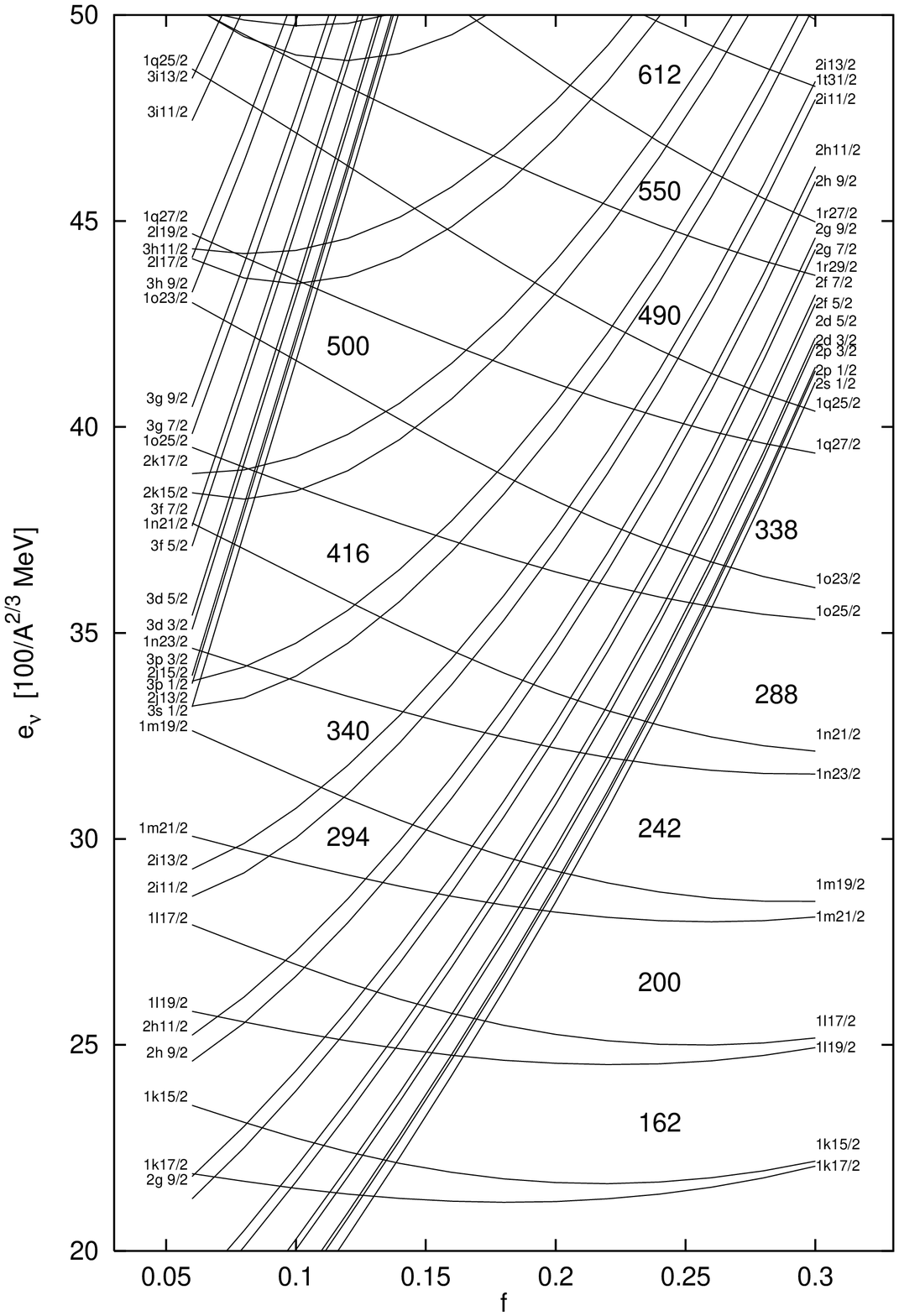}{180mm}{3}{}
\pagebreak[5]
\ifig{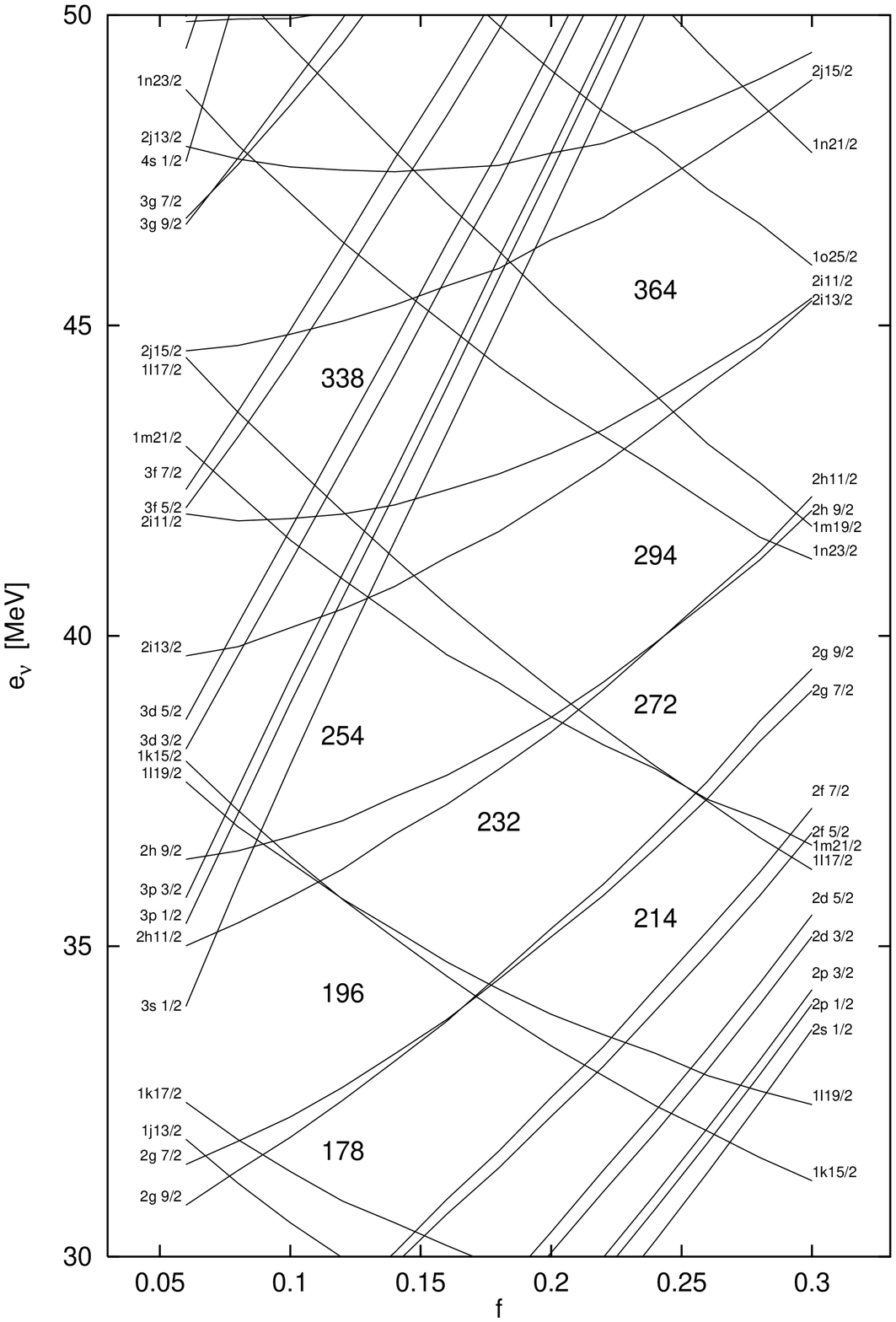}{180mm}{4}{}
\pagebreak[5]
\ifig{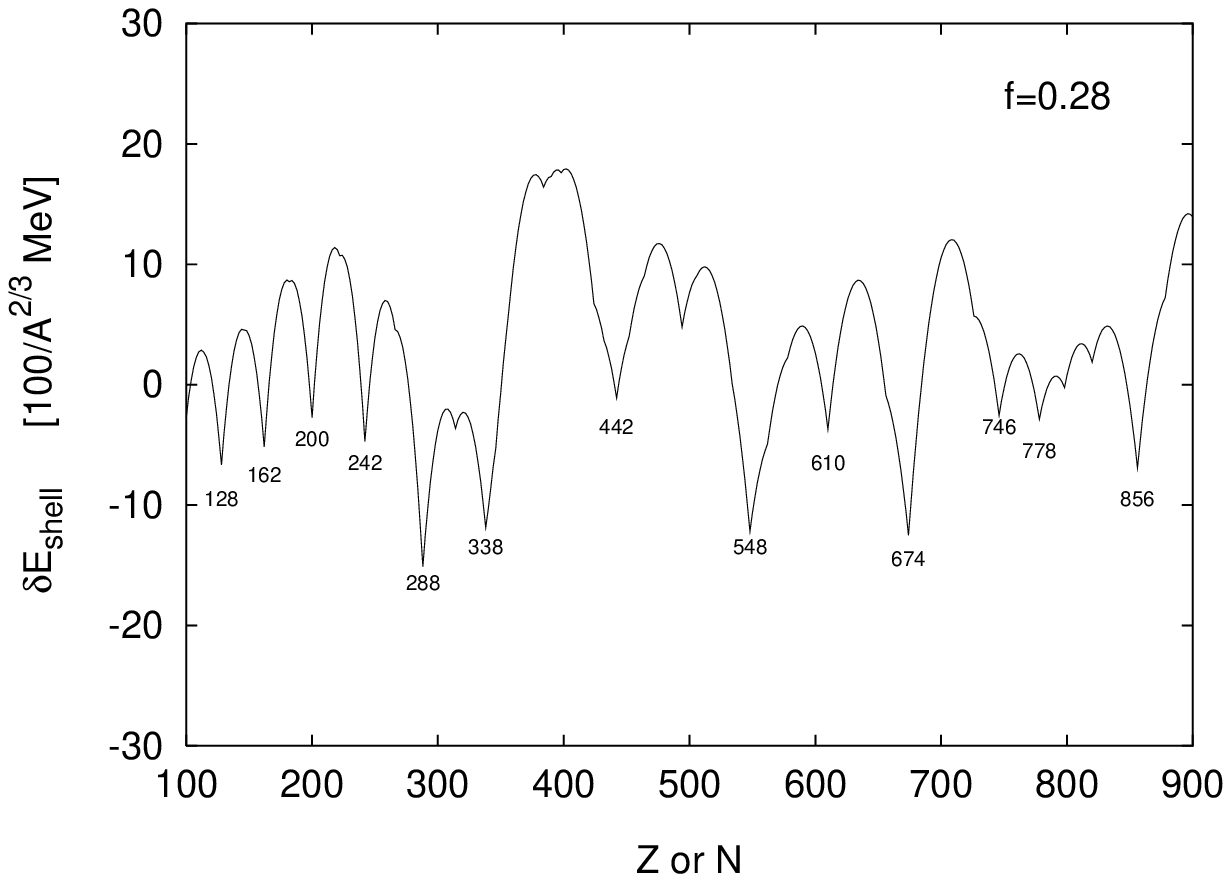}{100mm}{5}{}
\pagebreak[5]
\ifig{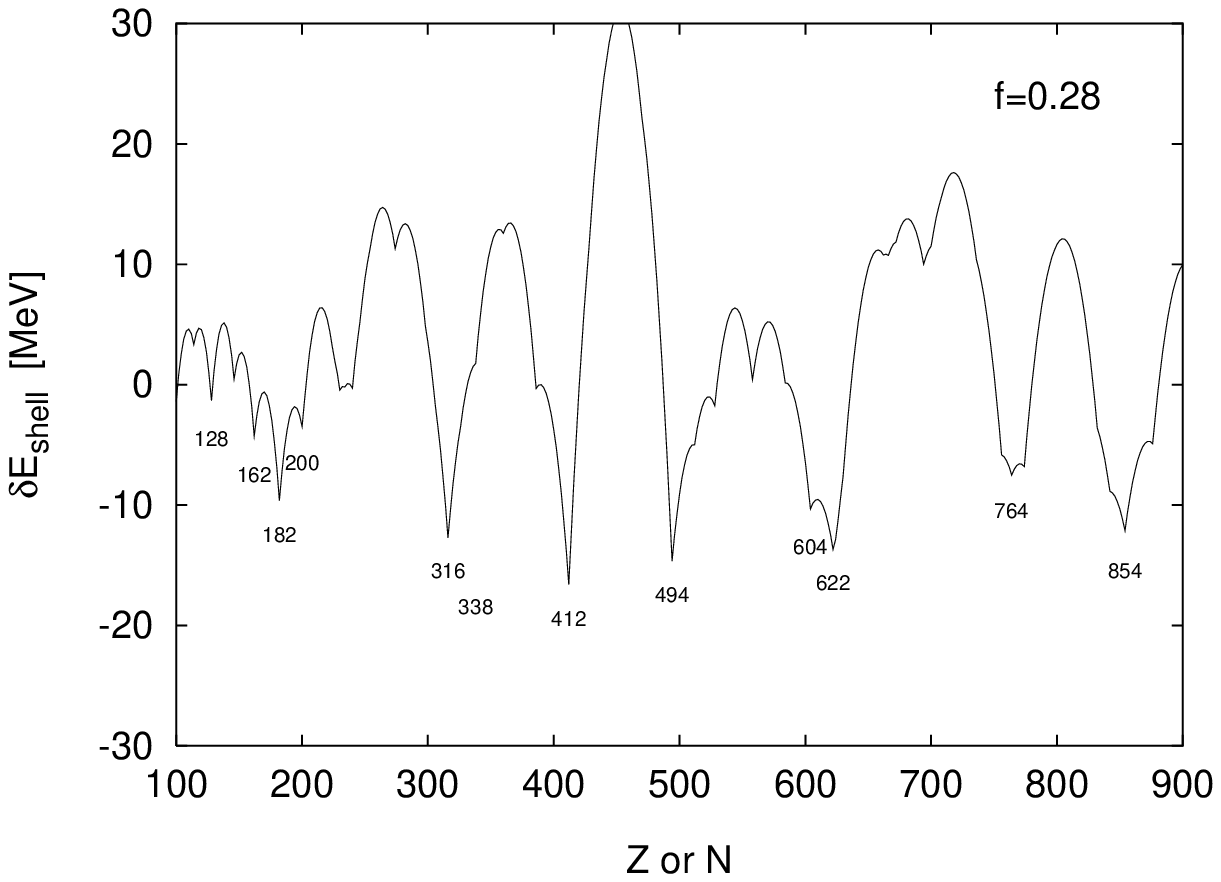}{100mm}{6}{}
\pagebreak[5]
\ifig{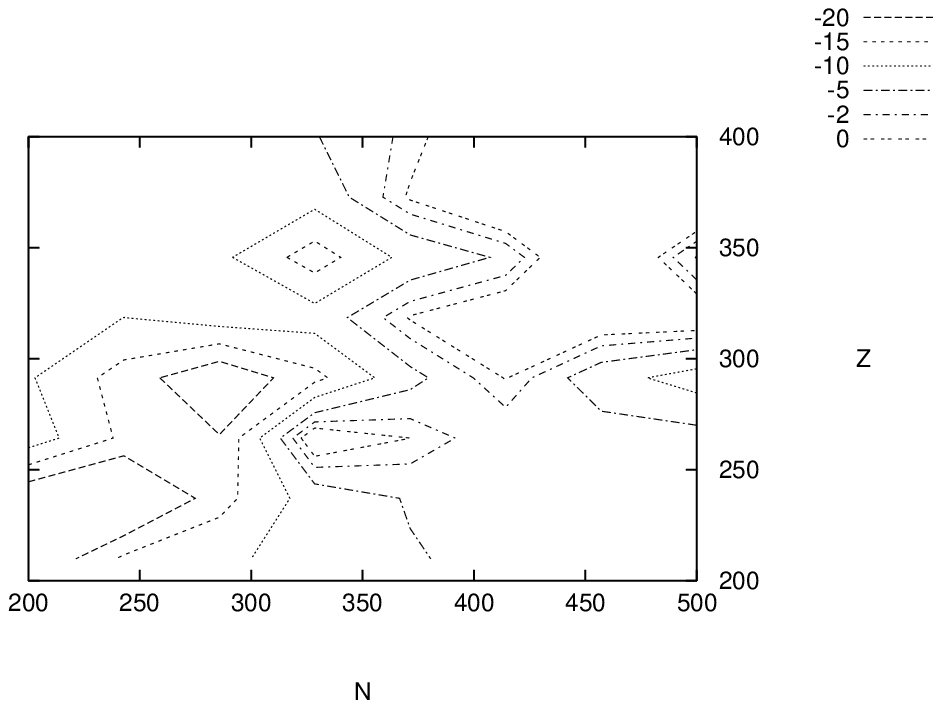}{100mm}{7}{}
\pagebreak[5]
\ifig{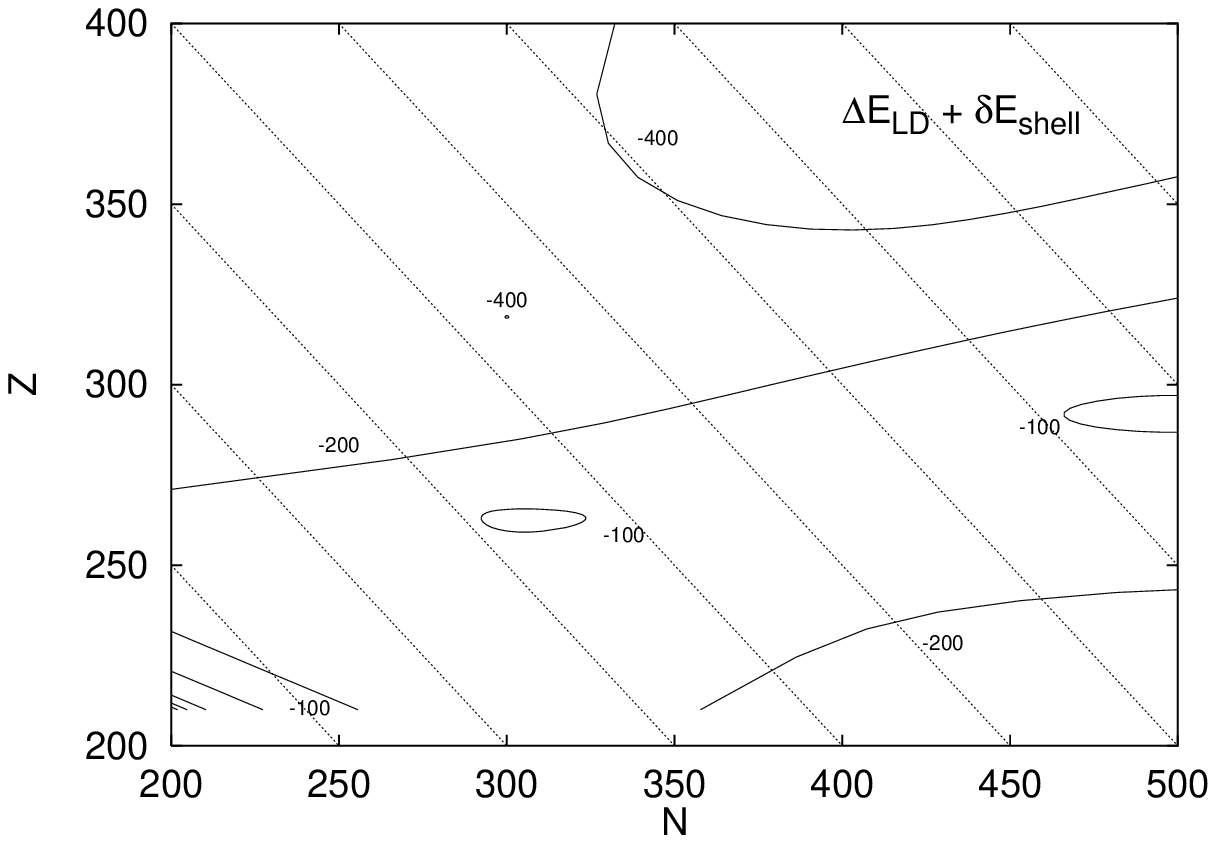}{100mm}{8}{}
\pagebreak[5]
\ifig{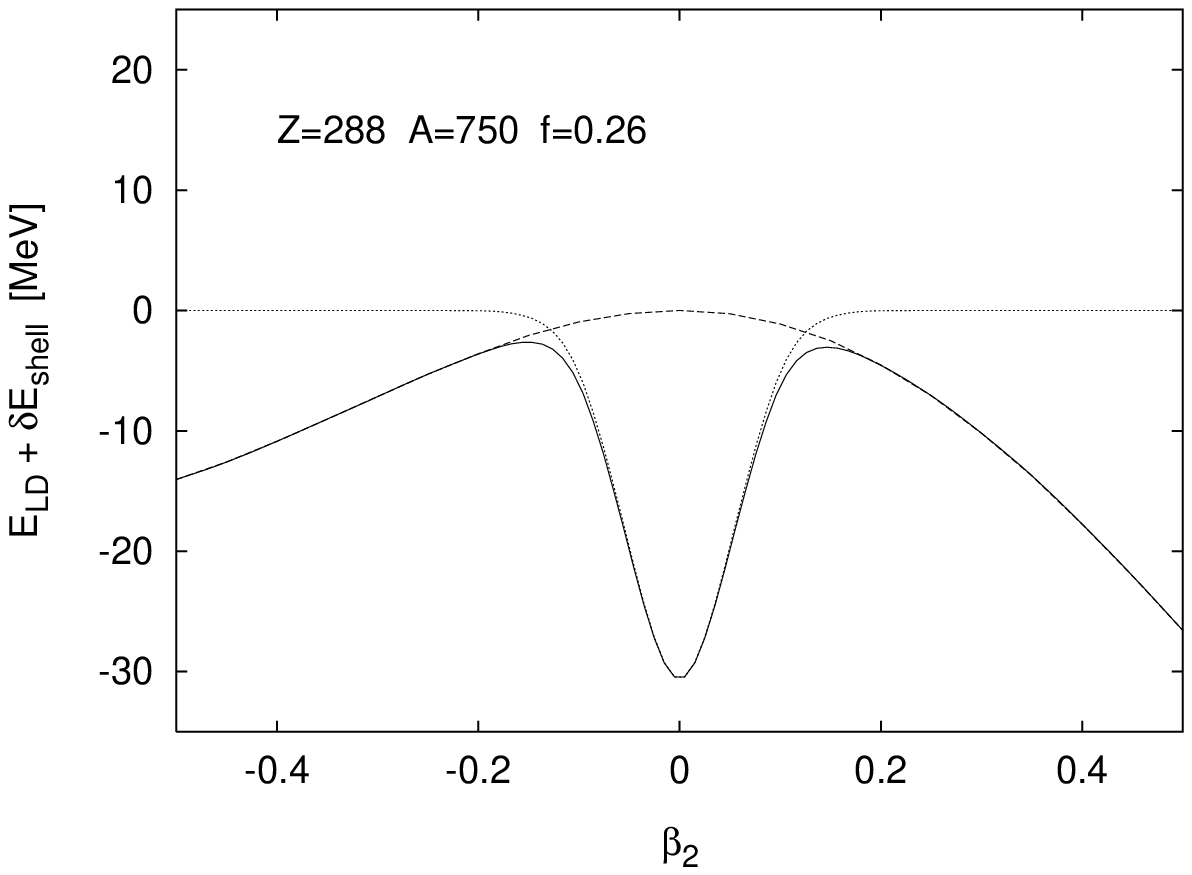}{100mm}{9}{}

\end{document}